\documentclass[a4paper, pra, aps, nopacs, twocolumn,nofootinbib]{revtex4-2}

\usepackage[utf8]{inputenc}  
\usepackage{graphicx} 
\usepackage{epsfig}
\usepackage[dvipsnames]{xcolor}
\usepackage{bm}        
\usepackage{dsfont}
\usepackage{amssymb}   
\usepackage{amsmath}
\usepackage{mathtools}
\usepackage{hyperref}
\hypersetup{colorlinks=true,citecolor=blue,linkcolor=red,
urlcolor=blue,pdfstartview=FitH,bookmarksopen=true}

\usepackage{comment}

\DeclareUnicodeCharacter{2010}{-}

\DeclareMathOperator{\tr}{tr}

\newcommand{\ket}[1]{\ensuremath{|#1\rangle}}

\newcommand{\braket}[2]{\ensuremath{\langle #1|#2\rangle}}
\newcommand{\ketbra}[1]{\ensuremath{| #1 \rangle \!\langle #1 |}}

\newcommand{\mean}[1]{\left\langle#1\right\rangle}
\newcommand{\id}{\ensuremath{\mathds{1}}}

\newcommand{\D}{\mathcal{D}}
\newcommand{\cH}{\mathcal{H}}
\newcommand{\cU}{\mathcal{U}}
\newcommand{\cV}{\mathcal{V}}

\def\one{{\mbox{$1 \hspace{-1.0mm}  {\bf l}$}}}

\newcommand{\eqnref}[1]{Eq.~(\ref{#1})}

\begin{document}

\title{Universal algorithms for quantum data learning}

\author{Marco Fanizza, Michalis Skotiniotis, John Calsamiglia, Ramon Mu\~noz-Tapia, and Gael Sent\'is}

\affiliation{
F\'isica Te\`orica: Informaci\'o i Fen\`omens Qu\`antics, Departament de F\'isica,
Universitat Aut\`onoma de Barcelona, 08193 Bellatera (Barcelona) Spain
}

\begin{abstract}
Operating quantum sensors and quantum computers would make data in the form of quantum states available for purely quantum processing, opening new avenues for studying physical processes and certifying quantum technologies. In this Perspective, we review a line of works dealing with measurements that reveal structural properties of quantum datasets given in the form of product states. These algorithms are universal, meaning that their performances do not depend on the reference frame in which the dataset is provided. Requiring the universality property implies a characterization of optimal measurements via group representation theory.
\end{abstract}

\maketitle

\section{Introduction}

Data analysis is a central element of the contemporary world, where data is abound but mostly featureless. The overall aim is to extract information from structured or unstructured data that generally comes from an unknown stochastic process. This information may be about the generating process or the data itself, e.g., a feature shared by all data instances or a global property of the analyzed dataset. 

For instance, a dataset $\mathcal{D}_C=\{x_i\}$ may allow us to infer some property of an unknown underlying distribution $p_X$ under the assumption of independent and identically distributed (iid) sampling; dropping this assumption, we may instead aim to identify some geometric  
property of $\mathcal{D}_C$; or if data points $x_i$ come with labels $y_i$, we may use the set $\{(x_i,y_i)\}$ to learn how to attach a label $y'$ to an unlabeled data point $x'$. The inference algorithms in these examples all try to learn some property of a given dataset, and their design is made with universality in mind: the analysis may only partially depend on certain generic features of the data, but not on how the data looks like in a particular instance, and therefore it works for any possible input that has these features.

In the emerging world of quantum technologies, one can conceive scenarios where not only the processing of information is done by exploiting quantum resources but also data itself is quantum. 
It is only natural to ask, what sort of information can we learn from a \emph{quantum dataset}, and how do we extract it?
In this Perspective, we review under a common framework a series of works that deal with these questions and share two important features: they are \emph{universal} algorithms in the above sense, and they use \emph{symmetries} present in the dataset to their advantage.

First, we need to define precisely what we mean by quantum data. We can think of a quantum particle in some state $\rho_i$ as the natural generalization of a sample $x_i$ from a classical probability distribution $p_X$, and hence a quantum dataset would be a collection of quantum states $\mathcal{D}_Q=\{\rho_i\}$. Importantly, we assume to have access to the quantum particles but not the density matrices representing their states, which are unknown to us. The states $\rho_i$ are the result of an inaccessible quantum process, in the same way as we do not control the generating processes of classical data samples $x_i$.

The types of properties that we may aspire to learn from a quantum dataset crucially depend on the assumed structure of $\D_Q$. For instance, if the data consists of $N$ identical copies of an unknown pure qubit state, e.g., $\D_Q = \rho^{\otimes N}$ with $\rho=\ketbra{\psi}$, we can try to learn the direction the state is pointing at; this is an \emph{individual} property of every element in $\D_Q$. However, if $\D_Q = \rho_0^{\otimes n}\otimes\rho_1^{\otimes N-n}$ with $\rho_0 \neq \rho_1$, the above property becomes ill-defined. Instead, if we know $n$, we may learn a \emph{relational} property of the components of the dataset, e.g., the overlap $\tr \rho_0\rho_1$. As a last example, in the case we do not even know $n$ in the dataset above, we can aim at learning it, that is, how to split $\D_Q$ into different subsets of identical copies. This would be a \emph{global} property of the set.

A powerful way to address this kind of problems is to consider the symmetries of $\D_Q$ under which the property of interest remains invariant. These symmetries are induced by the structure of the dataset, e.g., in the first example above, permuting the elements of $\D_Q$ does not alter the direction encoded in $\rho$. The symmetries of the properties determine that the optimal measurement can be chosen to respect the symmetries, simplifying both the design of the measurement and the analysis of the performance. 
Furthermore, measurements can be chosen to be sensitive only to changes in the property of interest and not depend explicitly on any other information contained in an instance of $\D_Q$, e.g., on the particular states $\rho_0$, $\rho_1$ for any of the problems outlined above. This type of quantum learning algorithms are thus \emph{universal}, agnostic to the input data as long as $\D_Q$ fulfils the symmetries prescribed by the problem at hand, and are consistent with the described notion of quantum data where the actual density matrix of $\D_Q$ is unknown to the experimenter.

Universal algorithms are generally optimized for two kinds of figures of merit: worst-case and average-case. In the worst-case approach, the interest is usually in quantifying the smallest dataset size (\textit{sample complexity}) such that the inference task is guaranteed to be successfully completed with high probability even with the worst possible input, and expressing this minimal size in terms of the extensive parameters of the problem, such as the dimension of the quantum systems involved~\cite{Montanaro2016}. In the average case approach, the interest is in quantifying the average performance 
 over all admissible
datasets of a given size, with a prior probability distribution over inputs reflecting the symmetries of the property of interest and our prior knowledge about the parameter we want to infer (e.g., \cite{Massar95}).\footnote{One could focus as well on worst-case performance over all possible inputs for fixed-size datasets, but this take is somewhat less frequent in the literature.} The latter approach
often allows us to work out explicit expressions for the optimal value of the figure of merit and to compute its asymptotics for large datasets, 
while the former usually reports only the asymptotic scaling of the sample complexity, but has the benefit of not having to specify any prior probability distribution.
Both approaches give interesting information about the performance of optimal tasks. While primitive tasks like tomography and identity testing are studied in depth in the worst-case scenario (for a review of these results, 
see~\cite{Kliesch2021}), problems with more structure such as supervised and unsupervised classification of quantum states and change point detection 
(described below in detail) have been analyzed mostly in the average-case asymptotic setting. This is also usually more natural since the task itself may not always be realizable with arbitrarily high probability of success as the size of the dataset increases, so that there is not an immediate notion of sample complexity. In this Perspective we want to highlight the importance of these structured problems as generalizations of learning tasks to quantum data, therefore we will see average-case results more in depth. It would be nonetheless interesting and 
relevant to formulate adequate worst-case questions for the same problems.

In order to illustrate the strength of symmetry principles in this context, we begin with the primitive example of estimating 
an unknown pure state given $N$ copies of it. We proceed with a more structured dataset in the case of programmable discrimination of pure qubits, followed by a formal description of the underlying techniques and the abstract general structure of the problem. We then review more complex results that share the same approach, and we end the paper with a discussion.

\section{A simple example}

State estimation~\cite{Massar95} is an essential primitive in quantum inference where given $n$ copies of an \emph{unknown state} $\ket{\psi}$ of dimension $d$ a suitable measurement strategy is devised in order to infer the state of the system. Given a measurement characterized by a positive operator-valued measure (POVM) 
$E=\{E_{\alpha}\geq 0, \sum_{\alpha}E_{\alpha}=\id\}$,  an estimate $\ket{\phi_\alpha}$ needs to be produced for each of the outcomes $\alpha$ such that it resembles as much as possible the input state $\ket{\psi}$, as measured by the fidelity $f(\phi_{\alpha},\psi)=|\braket{\phi_{\alpha}}{\psi}|^{2}$. In such a setting it is quite natural to designate a prior distribution for our dataset:  assuming that all pure states $\ket{\psi}$ are equally likely singles out the unitarily invariant measure $d\psi$. Once equipped with a prior distribution of the data we can assess the performance of a given strategy by the average fidelity over all input states and all measurement outputs. The optimal strategy is given by the following Bayesian optimization:
\begin{equation}
F_{\mathrm{ave}}:=\mean{f}=\max_{E} \int d\psi\sum_{\alpha}f(\phi_{\alpha},\psi) \tr\left(E_{\alpha}\ketbra{\psi}^{\otimes n}\right) .
\label{eq:aFid}
\end{equation}
We note here that in general there is no established criteria to designate a prior distribution on the input data ---this is the main caveat of such Bayesian formulations. 
Alternatively, one can define a worst-case figure of merit for $F_{\mathrm{wc}}= \min_{\psi} \max_{E} \sum_{\alpha}f(\phi_{\alpha},\psi)\tr\left(E_{\alpha}\ketbra{\psi}^{\otimes n}\right)$, which does not require to fix a prior.

Next we show how the symmetries readily lead to elegant solutions to the optimization problem. From Eq.~\eqref{eq:aFid} and writing $\psi=\ketbra{\psi}$,
\begin{align}
F_{\mathrm{ave}} &=\max_{E} \int\!d\psi\sum_{\alpha}\tr({\psi}\phi_{\alpha})\tr\left(E_{\alpha}\psi^{\otimes n}\right)\\
&= \max_{E} \int\!d\psi\sum_{\alpha}\tr\left[(\phi_{\alpha}\otimes E_{\alpha}){\psi}^{\otimes n+1}\right]\\
&=\frac{1}{d_{n+1}}\max_{E} \sum_{\alpha}\tr\left[(\phi_{\alpha}\otimes E_{\alpha})\id_{\mathrm{sym}}^{(n+1)}\right] \label{eq:sym}\\
&\leq \frac{1}{d_{n+1}}\max_{E} \sum_{\alpha}\tr\left[(\phi_{\alpha}\otimes E_{\alpha})\right]\label{eq:ineqEstimation}\\
&=\frac{1}{d_{n+1}}\tr(\sum_{\alpha} E_{\alpha})= \frac{d_{n}}{d_{n+1}}=\frac{n+1}{n+d} \,, \label{eq:comple}
\end{align}
where in \eqref{eq:sym} we have used that  $\int\!d\psi \psi^{\otimes n}$ is  the unique trace one operator with support in the fully symmetric subspace\footnote{The fully symmetric subspace of $n$ qudits is $\mathrm{span}\{ \ket{\Psi^{(n)}}: P_{\sigma}\ket{\Psi^{(n)}}=\ket{\Psi^{(n)}} \mbox{for all permutations $\sigma$ over $n$ elements} \}$.} of $n$ qudits that is invariant under rigid $U^{\otimes n}$ unitary transformations, i.e., the projector $\id_{\mathrm{sym}}^{(n)}$ normalized by its dimension $d_{n}=\binom{n+d+1}{n}$ (see Schur lemma \cite{hayashi2017group1}); in \eqref{eq:comple} we have also used that the input data lies on the symmetric subspace and hence w.l.o.g. we can take $\sum_{\alpha} E_{\alpha}=\id_{\mathrm{sym}}^{(n)}$.

On the other hand it is straightforward to check that the inequality in \eqref{eq:ineqEstimation} can be attained by picking a covariant continuous POVM  with elements $E_{\phi}=d_{n} \phi^{\otimes n}$,  which fulfill $\int \! d\phi E_{\phi}=\id_{\mathrm{sym}}^{(n)}$, and a corresponding estimate $\phi$. 
Note that this optimal covariant strategy returns an equally good estimate for each possible input state, hence the average fidelity coincides with the worst case fidelity $F_{\mathrm{wc}}$.

\section{Programmable discriminators}

Programmable discriminators are universal devices capable of performing binary classification of an unknown quantum state when the information about the two possible classes is provided as quantum data. This is in contrast to the standard state discrimination scenario, where we are provided with the classical description of the possible states of the system. In their simplest form, they take as input a quantum dataset of the form $\D_Q=\ket{\psi_0}_A\otimes\ket{\psi_i}_B\otimes\ket{\psi_1}_C$, where $i=0,1$ and $\ket{\psi_{0,1}}$ are unknown qubit states. Registers $A,C$ are termed \emph{program ports}, and take the two different ``template'' quantum states. 
The register $B$ is called \emph{data port}, and the goal of the device is to identify the value of the label $i$.

We demand such device to be universal, i.e., that it works for any pair of program states $\ket{\psi_{0,1}}$. 
Choosing an average case approach,
we say that a programmable device performs optimally if  the average error rate in the label identification with respect to a uniform distribution of pairs of states is the minimal one.  
The key observation that exploits the symmetries of $\D_Q$ and allows to solve the problem comes from the following consideration:
if the data state is $\ket{\psi_0}$, the effective state entering the device is
\begin{eqnarray}
\label{sigma0}
\sigma_0  = 
\int d\psi_0 d\psi_1  \psi_0^{\otimes 2}\otimes 
\psi_1 = \frac{1}{6}\id^{AB}_{\rm sym}\otimes\id^C \,,
\end{eqnarray}
where we have applied Schur lemma~\cite{hayashi2017group1} as in deriving Eq.~\eqref{eq:sym}, and we have used that $d_2 d_1 =6$ for qubits.
Here $\id^{AB}_{\rm sym}$ 
is the projector onto the fully symmetric space of qubit systems $A$ and $B$, and 
$\id^C$ is the identity matrix in the qubit system $C$. 
Likewise, if the data state is $\ket{\psi_1}$ one has 
\begin{equation}
\sigma_1= \frac{1}{6}\id^A\otimes \id^{BC}_{\rm sym} \,.
\end{equation}

Notice that the spectrum of both matrices $\sigma_0$ and $\sigma_1$ is identical and that the basis elements of their support simply differ in the way the three qubits are coupled, hence this information is the only one that can be used to distinguish the two effective states. The states are diagonal in the total angular momentum basis $\ket{JM;q_i}$, where $q_i$ tags the two different ways the qubits are coupled, i.e., $q_0=\{AB, ABC\}$ and $q_1=\{BC, ABC\}$, to arrive to $J=3/2,1/2$.
Furthermore, we note that the overlap between the two bases fulfils
$
\braket{JM;q_0}{J'M';q_1}=C_J \delta_{JJ'}\delta_{M M'} \,,
$
that is, 
the relevant quantum numbers for distinguishing $\sigma_0$ from $\sigma_1$ are $J$ and $q_i$. The numbers $C_J$ are called Racah coefficients~\cite{Messiah14}, which in this example take the values $C_{3/2} = 1$ and $C_{1/2}=1/2$.

The optimal classification procedure is then provided by first measuring $J$,
and, upon having obtained a specific result with probability $p_J=\mu_J/6$, with $\mu_J$ the dimension of the subspace $J$, a subsequent Helstrom measurement~\cite{Helstrom1976} that optimally distinguishes the pure states $\ket{JM,q_0}$ and  $\ket{JM,q_1}$ and has error probability $P_e^J=[1-\sqrt{1-C_J^2}]/2$.
The overall minimum error probability reads
\begin{align}
\label{Pe}
P_e =
\sum_{J=1/2}^{3/2} p_{J} P_e^J
 =\frac{1}{2}\left(  1-\frac{1}{2\sqrt{3}}\right) \,.
\end{align}
In contrast to the previous example of state estimation, where the covariant continuous POVM is sensitive to the quantum number $M$ encoding information about the direction where $\ket{\psi}$ is pointing at, here this information is irrelevant.
Observe that Eq.~\eqref{Pe} is solely a function of invariant quantities such as dimensions and overlaps $C_J$.
In spite of the little information we had about the states (not even its classical description) only
roughly one third of the times (on average) the data state will be wrongly classified.

\section{The general framework}

The previous two examples demonstrate how a clever use of the inherent symmetries 
possessed by the quantum data sample leads to elegant, closed form expressions for 
assessing the average performance of the requisite task.  We now introduce the 
mathematical framework that such symmetries impose on the state space of a quantum 
dataset, and show how such framework can be generally exploited to significantly constraint the 
search for the optimal measurement~\cite{hayashi2017group2}.

Let $\cH_d$ be a $d$-dimensional  Hilbert space, and let us assume that $\D_Q$ is an element of the space of linear operators over the $n$-fold tensor product of $\cH_d$, $\D_Q\in\mathcal{L}(\cH_d^{\otimes N})$.
The tensor product space $\cH_d^{\otimes N}$ naturally carries the action of two fundamental symmetry 
groups:  that of the {\it special unitary group} in $d$ dimensions $\mathrm{SU}(d)$, as well as that of the {\it permutation 
group} of $N$ elements, $\mathrm{S}_N$.  The action of these two groups  on any state $\D_Q$ is mediated through 
their respective {\it unitary representations}, $\{U^{\otimes N}_g\in \mathrm{GL}(d), \forall\, g\in \mathrm{SU}(d)\}$, 
$\{V_\tau\in \mathrm{GL}(d), \forall\, \tau\in \mathrm{S}_N\}$, whose effect on the orthonormal basis 
$\{\ket{i_1,\ldots i_N}:=\ket{i_1}\otimes\ldots\otimes\ket{i_N}\}$ of $\cH^{\otimes N}_d$ is given by 
	\begin{equation}
		\begin{split}
			U^{\otimes N}_g\ket{i_1,\ldots i_N}&= U_g\ket{i_1}\otimes\ldots\otimes U_g\ket{i_N}   \\
			V_\tau\ket{i_1,\ldots i_N}&=\ket{i_{\tau^{-1}(1)}}\otimes\ldots\ket{i_{\tau^{-1}(N)}}\,    .
		\end{split}
		\label{eq:reps}
	\end{equation}

Using Schur lemmas~\cite{hayashi2017group1}, the representations $U_g$ and $V_\tau$ in \eqnref{eq:reps} can be decomposed into a direct sum of
{\it irreducible representations} (irreps), $U_g^{(\lambda)},\, V_\tau^{(y)}$, as follows
	\begin{equation}
		\begin{split}
		U^{\otimes N}_g&= \bigoplus_{\lambda} \, U^{(\lambda)}_g\otimes \one^{(\lambda)}\quad \forall\, g\in \mathrm{SU}(d)\,,\\
		V_\tau&=\bigoplus_{y}\, V^{(y)}_\tau\otimes \one^{(y)}     \quad \forall\, \tau\in\mathrm{S}_N\,,
		\end{split}
		\label{eq:Schur}
	\end{equation}
where $\lambda, y$ label {\it inequivalent} irreps of $\mathrm{SU}(d)$ and $\mathrm{S}_N$, and 
$\one^{(\lambda)}, \one^{(y)}$ have dimensions $\nu_\lambda, \mu_y$ corresponding to the number of times 
the irreps $U_g^{(\lambda)},\, V_\tau^{(y)}$ appear in the decomposition of the corresponding representations.  
For the symmetric group the irrep label $y$
denotes the ordered, integer partitions of $N$ into at most $d$ parts, i.e., 
	\begin{equation}
		y:=\{(y_1,\ldots , y_d)\, \vert \, y_1\geq \ldots\geq y_d\, \mathrm{and}\, \sum_{k=1}^d y_k=N\}\, .
		\label{eq:partitions}
	\end{equation}
The irrep label $\lambda$ of $\mathrm{SU}(d)$, 
on the  other hand, has a rather more complicated interpretation.\footnote{For the interested reader, the irrep label 
$\lambda$ is related to the highest weight of the associated Lie algebra~\cite{goodman2009symmetry}.}  For the case where $d=2$, 
$\lambda$ labels the {\it total angular momentum} quantum number.  

The irrep labels $\lambda$ and $y$ appear, at first sight, unrelated, but the irreps with non-zero multiplicity in the decomposition are in fact in one-to-one correspondence due to an 
important result in representation theory known as Schur-Weyl duality~\cite{goodman2009symmetry}.
Noting that $[U^{\otimes N}_g, 
V_\tau]=0,\, \forall\, g\in\mathrm{SU}(d), \, \mathrm{and}\, \forall\, \tau\in\mathrm{S}_N$, implies that the decomposition of 
$U_g^{\otimes N}$ in \eqnref{eq:Schur} is block-diagonal with respect to the irrep decomposition of $V_\tau$ and {\it vice versa}.  
Schur-Weyl duality tells us
that one may use the labels $y$
to decompose {\it both} representations, where only products of irreps with the same labels appear in the decomposition.
For 
$\mathrm{SU}(2)$, $\lambda$ labels the values of the total angular momentum $J$ of $N$ spin-$\frac{1}{2}$ particles
and can be obtained from the corresponding integer partitions as $\lambda=\frac{y_1-y_2}{2}$.

The block decomposition of \eqnref{eq:Schur} induces a similar decomposition on the state space $\cH_d^{\otimes N}$ as
	\begin{equation}
		\cH_d^{\otimes N}\cong \sum_y \cH^{(y)}:=\sum_y \cU^{(y)}\otimes \cV^{(y)}\, ,
		\label{eq:Schur-Weyl2}
	\end{equation}
where we have used the irrep labels of $\mathrm{S}_N$ to label the various {\it invariant subspaces} $\cH^{(y)}$.  
The latter can be further decomposed into a 
tensor product of subspaces $\cU^{(y)},\, \cV^{(y)}$ which support the 
irreps $U_g^{(y)}$ and $V_\tau^{(y)}$ respectively.  The congruence sign in Eq.~\eqref{eq:Schur-Weyl2} indicates that there exists an 
orthonormal basis relative to which the total state space $\cH_d^{\otimes N}$ assumes the specific block decomposition. This change
of basis is accomplished by the {\it Schur transform} for which efficient circuits 
that implement it
exist~\cite{Bacon2006,Krovi2019}.

It is worthwhile pausing briefly to understand the nature and role of the block-diagonal decomposition in \eqnref{eq:Schur-Weyl2}. 
The irrep labels $\lambda,\, y$ encodes {\it globally invariant} properties of the dataset, an example of which is the 
spectrum of eigenvalues of any $\rho$ when $\D_Q=\rho^{\otimes N}$. Indeed, the vector $y/N$ concentrates on the spectrum of $\rho$~\cite{alicki1988symmetry,Keyl2001,hayashi2002quantum,christandl2006spectra}.
In general, the subspace $\cU^{(y)}$ encodes information pertaining to properties associated with the symmetry group $\mathrm{SU}(d)$, 
such as the expectation value of some generator of $\mathrm{SU}(d)$,\footnote{For $d=2$ for instance, $\cU^{(y)}$ encodes information about the total angular momentum 
about some axis $\hat{\bf n}$.} which are \emph{invariant} under all permutations. Conversely, the subspace $\cV^{(y)}$ encodes
information associated with the permutation group $S_N$---such as the labels of the constituent particles---which are in turn 
invariant under arbitrary rotations.  

A case of particular relevance is when every admissible quantum dataset is {\it invariant} with respect to a group of 
transformations, i.e., $\mathrm{Inv}:=\{\D_Q^{(i)}\, \vert W_g\,\D^{(i)}_Q\, W_g^\dagger=\D^{(i)}_Q, \, \forall \, g\in G, \mathrm{and}\, \forall\, i\}$. In this case it is sufficient to consider invariant measurements~\cite{hayashi2017group2}, i.e., 
	\begin{equation}
		M^{\mathrm{inv}}:=\{M_i\, \vert\, W_g\, M_i\, W^\dagger_g=M_i, \forall\, g\in G\}.
		\label{eq:invariantPOVM}
	\end{equation}
Using \eqref{eq:Schur-Weyl2} every element, $M_i$, of such a POVM can be written in block diagonal form.  For instance, if $G=\mathrm{SU}(d)$ then
	\begin{equation}
		M_i\cong \sum_y \one^{(y)}\otimes M_i^{(y)}\, .
		\label{eq:sudecominvariantPOVM}
	\end{equation}

A second case of particular importance is that of learning {\it covariant} properties of admissible quantum datasets, i.e., $\mathrm{Cov}:=\{\D_Q^{(g)}\,\vert\, \D_Q^{(g)}=W_g \D_Q W_g^\dagger, \, g\in G\}$ for some fiducial dataset $\D_Q$.  Notice that learning a covariant property reduces to learning the group element $g\in G$.
In this case it is sufficient to consider covariant measurements~\cite{hayashi2017group2}:  
	\begin{equation}
		M^{\mathrm{cov}}:=\{M_g \,\vert\, W_g\, M_0\, W^\dagger_g = M_g, \forall\, g\in G\}.
		\label{eq:cov} 
	\end{equation}
Since $\hspace{2mm}\mathclap{\displaystyle\int}\mathclap{\textstyle\sum}\,\,\, M_g\, \mathrm{d}g=\one$, where $\mathrm{d}g$ denotes the {\it 
invariant (Haar) measure} of $G$, Schur-Weyl duality imposes a block diagonal structure on the fiducial POVM element $M_0$.
If $G=\mathrm{SU}(d)$ then  
	\begin{equation}
		M_0\cong \sum_y M_0^{(y)},
		\label{eq:sudecompPOVM}
	\end{equation}
where $M_0^{(y)}\in\mathcal{L}(\cH^{(y)})$ and $\tr_{\cU^{(y)}}M_0^{(y)}=\one^{(y)}\in\mathcal{L}(\cV^{(y)})$.  Similarly, if $G=S_N$, 
$\tr_{\cV^{(y)}}M_0^{(y)}=\one^{(y)}\in\mathcal{L}(\cU^{(y)})$. 

We note that both covariant and invariant measurement strategies are known to be optimal both  when the figure of merit is chosen to be the average and the worst case~\cite{hayashi2017group2}.  We also note that in hybrid cases where the admissible quantum datasets are covariant but the property of interest, $f$, is invariant, i.e., $f(\D_Q)=f(W_g\D_Q W_g^\dagger), \forall\, g\in G$, then the optimal measurement is invariant.

\section{Further applications}

We are now ready to review a selection of more complex structured quantum learning problems that can be addressed with the methods we have presented. They have a common structure in that the relevant information is extracted from the distribution of the irrep label $y$ in the global Schur-Weyl decomposition and, possibly, from a measurement which is non-trivial only on the factor $\mathcal V^{(y)}$. 
Specifically, for a quantum dataset
\begin{equation}
    \D_Q=\sigma=\bigotimes_{i=1}^l\rho_i^{\otimes n_i}
\end{equation}
we can apply Schur-Weyl decomposition to each iid part $\rho_i^{\otimes n_i}$, $\sum_{i=1}^l n_i=n$ and obtain
\begin{equation}
\sigma=\bigotimes_{i=1}^l\sum_{y_i\vdash n_i}\pi^{(y_i)}(\rho_i)_i\otimes \one_i^{(y_i)},
\label{eq:sigma}
\end{equation}
where $y_i\vdash n_i$ denotes the various integer partitions of $n_i$, and $\pi^{(y_i)}(\rho_i)_i$ is a positive semi-definite operator 
that depends on $\rho_i$.\footnote{More precisely, $\pi^{(y_i)}$ is a representation of the general linear group in $d$ dimensions.} 
According to the same decomposition, we can write
\begin{align}
\mathcal H_d^{\otimes n}&=\bigotimes_{i=1}^l\bigoplus_{y_i\vdash n_i}\mathcal U^{(y_i)}\otimes \mathcal V^{(y_i)}=\bigoplus_{\vec{y}\in\mathfrak{S}}\bigotimes_{i=1}^l\mathcal U^{(y_i)}\otimes \mathcal V^{(y_i)}\nonumber\\
&=\bigoplus_{\vec{y}\in\mathfrak{S}}\bigoplus_{y\vdash n}\cU^{(y)}\otimes \mathcal W_{\vec{y};y}\otimes_{i=1}^l 
\mathcal V^{(y_i)}\nonumber\\
&=\bigoplus_{y\vdash n} \mathcal U^{(y)}\otimes \left(\bigoplus_{\vec{y}\in\mathfrak{S}} \mathcal W_{\vec{y};y}\otimes_{i=1}^l \mathcal V^{(y_i)}\right) ,
\label{eq:Schur-Marco}
\end{align}
where $\mathfrak{S}:=\lbrace \vec{y}=(y_1,\ldots, y_l)\, \vert y_1\vdash n_1,\ldots, y_l\vdash n_l\rbrace$, $\cU^{(y)}$ arises from the decomposition of $\otimes_{i=1}^l \cU^{(y_i)}$ into irreps with $\mathcal{W}_{\vec{y};y}$ the corresponding multiplicity space of this decomposition (possibly zero-dimensional). Note that the second factor in \eqnref{eq:Schur-Marco} is isomorphic to the space $\cV^{(y)}$ in \eqnref{eq:Schur-Weyl2}. Under an $\mathrm{SU}(d)$-invariant measurement, the state $\sigma$ of \eqnref{eq:sigma} produces the same statistics over measurement outcomes as an $\mathrm{SU}(d)$-invariant state of the form
\begin{equation}\label{eq:sigma-tilde}
\tilde{\sigma}=\bigoplus_{y\vdash n} \frac{\one^{(y)}}{\mu_y}\otimes \left(\bigoplus_{\vec{y}\in\mathfrak{S}} \xi_{\vec{y};y}(\rho_{1},\ldots,\rho_l)\bigotimes_{i=1}^l \frac{\one^{(y_i)}}{\nu_{y_i}}\right),
\end{equation}
where all the information is contained within the positive semi-definite operators  $\xi_{\vec{y};y}(\rho_{1},\ldots,\rho_l)$. 

\subsection{Programmable state discrimination}
We already solved this problem in a simple case. In a more general scenario, 
the problem consists in guessing the value of $i=a,b$ in a quantum dataset \mbox{$\sigma_i = \rho_1^{\otimes n_1}\otimes \rho_{2,i}^{\otimes m}\otimes \rho_3^{\otimes n_3}$} (hence a \emph{global} property of the set), where we know $n_1,\,n_3,\,m$, $\rho_1,\,\rho_3$ are unknown states and $\rho_{2,a}=\rho_1, \rho_{2,b}=\rho_3$. An optimal measurement will be a projection in the subspaces $(y_1,y_2,y_3);y$, followed by a binary measurement to distinguish between the two possible operators $\xi_{\vec y;y}(\rho_1,\rho_{2,i},\rho_3)$. Early works in programmable discriminators are \cite{Sasaki2002,Hayashi2006,He2007}.
The problem as stated here was solved for the average error figure of merit for arbitrary pure and mixed qubit states, including explicit asymptotics, in \cite{Sentis2010,Fanizza2018}.
When $\rho_1$ and $\rho_3$ are pure and their overlap is known, asymptotics of the optimal performance for $n_1,\,n_3,\,m\to\infty$ were obtained in \cite{Akimoto2011}, and for $m=1$ in \cite{Fanizza2018}.
There, it was observed that the optimal binary measurement in each subspace 
$(y_1,\, y_2,\,y_3);y$ is unique and independent of the overlap between the quantum states. 
Furthermore, the case of pure input states has also been considered in an unambiguous setting, where no error is allowed in making the guess for $i$ (see, e.g., 
\cite{Bergou2005,Sentis2010}).

\subsection{Supervised quantum learning}
Programmable discrimination can be viewed as a supervised quantum learning task where the pre-classified program states $\rho_1^{\otimes n_1}$ and $\rho_3^{\otimes n_3}$ comprise a quantum training dataset, and the classification of the test states $\rho_{2,i}$ is done using the information gained during training. This viewpoint arises naturally when considering training and classification as two separate phases of the protocol, which leads to imposing locality constraints on the optimal measurement (states $\rho_1^{\otimes n_1}\otimes\rho_3^{\otimes n_3}$ are measured first, then, conditioned on the outcome obtained, $\rho_{2,i}^{\otimes m}$ are measured and classified accordingly). For $m=1$, \cite{Sentis2012a} showed that such split training-and-testing strategies are optimal even for finite training data, and that estimating $\rho_{1}$ and  $\rho_{3}$ separately and performing the corresponding Helstrom measurement on the test is suboptimal. The latter was also found to be suboptimal for a worst-case figure of merit \cite{Guta2010}. 
Furthermore, the same question was analyzed when the input states are coherent states \cite{Sentis2014a,Zoratti2021} and a gap was shown to exist. The difference between split and global strategies was considered more generally in \cite{Monras2017} for generic supervised quantum learning problems. In this work, the gap was proven to vanish for average figures of merit when $m\to\infty$, regardless of the structure of the training dataset.

\subsection{Unsupervised classification of quantum data}
This problem was analyzed in \cite{Sentis2019,Spencer-Wood2022} in its binary form. 
The quantum dataset is a product state $\sigma_{\vec x}=\otimes_{i=1}^n \rho_{x_i}$, where $x_i=0,1$, the index binary vector $\vec x$ specifies the arrangement of the two types of states $\rho_0,\,\rho_1$ which are pure and unknown, and we want to infer with minimum average error probability a \emph{clustering} of $\sigma_{\vec x}$ into two subsets of identical states, a task reminiscent to clustering protocols in classical machine learning which find subsets of points generated from the same probability distribution. In this problem, learning a clustering amounts to learning the number of states initialized as $\rho_0$, the number of states initialized as $\rho_1$, and their positions. 
Note that $\sigma_{\vec x}$ can again be considered to have the form in Eq.~\eqref{eq:sigma-tilde}, but the sum over $\vec y$ now only has one term. The optimal measurement strategy consists in projecting onto the subspaces $y$, and then distinguishing among all the possible operators $\xi_y(\vec x)$. Despite there being $2^{n-1}$ possible input states $\sigma_{\vec x}$, in \cite{Sentis2019} it is shown that the optimal success probability of this task scales only as $O(1/n)$, and, remarkably, that it is an increasing function of the local dimension $d$. In \cite{Spencer-Wood2022}, optimal measurement strategies were analyzed for the same task when $n=3$, and a trade-off was quantified between classifying the first two states and subsequently the third versus directly clustering the three states.

\subsection{Quantum change point detection}
Change point detection is a very active area in statistical inference \cite{Tartakovsky2020} where the task is to identify 
 the moment when the underlying probability distribution of a monitored stochastic process changes. Edge detection is the analogous problem where the change occurs in space rather than in time. 
These primitives find applications in a variety of contexts, including quality control, navigation, biology, or finance, \cite{Tartakovsky2020} and are generally useful in any problem involving the analysis of a sequence of samples, since this requires the stability of system parameters. 
The problem has  been  tackled only recently in the quantum setting. In its simplest instance a source emits iid states $\rho_0$ until point $k$ when it starts producing another state $\rho_1$, where both $\rho_0,\, \rho_1$ are pure states.  The problem is then to determine the change point $k$ by suitable measurement on the states $\{\rho_0^{\otimes(k-1)}\!\otimes\!\rho_1^{\otimes(n-k+1)}\}$. A closed expression for the optimal  probability of success was given in \cite{Sentis2016} under the most general collective measurement scheme, and in \cite{Sentis2017} for the unambiguous identification. In both cases the success probability can be seen to approach a constant as the length of the sequence, and hence the number of hypothesis, increases.  The techniques introduced here allow one  to extend the study to the very relevant setting where either $\rho_1$, or both are unknown \cite{inpreparation}. Each hypothesis $k$ is fully characterized by $\sigma_{k}=\rho_0^{\otimes (k-1) }\otimes \id^{\mathrm{sym}}_{k\ldots n}$ in the known-to-unknown case and 
 $\tilde{\sigma}_{k}=\id^{\mathrm{sym}}_{1\ldots k-1}\otimes \id^{\mathrm{sym}}_{k\ldots n}$ in the unknown-to-unknown case. In either case all the relevant information resides in a rank-one, unnormalized state $\psi^{(k)}$ in each of the relevant permutation group irreps: i.e., $\sigma_{k}=\sum_{y} \ketbra{y, m}\otimes \tilde\psi_{y,m}^{(k)},\, \tilde{\sigma}_{k}=\sum_{y} \id_{y}\otimes \tilde \psi_{y}^{(k)}$ respectively. Note that the former case has full SU(d) invariance while the latter is invariant only under the subgroup which leaves the initial state invariant 
 $U\rho_0 U^{\dagger}=\rho_0$.\footnote{The quantum number $m$ corresponds to the observable $Z=\sum_{k=0}^{n}\id^{\bar 0}\otimes\stackrel{k}{\ldots}\otimes\rho_0\otimes\id^{\bar 0}{\ldots}$ labeling the 1-d irreps of that subgroup.} The problem hence reduces to discriminating among $n$ pure quantum states for each block $y$. Quite surprisingly, for long sequences of quantum states the success probability in these more adverse situations, with unknown states, attains the same asymptotic value as the average obtained for known states \cite{inpreparation}.

\subsection{Learning of distance measures}
Finally, let us mention another important type of property of $\D_Q$ we may wish to learn: \emph{relational} properties among the $l$ states of a set $\mathcal S=\{\rho_1,...,\rho_l\}$.  Such 
properties are captured by unitarily invariant 
quantifiers, i.e., 
$h(\rho_i,...,\rho_l)=h(U\rho_i U^{\dagger},...,U \rho_l U^{\dagger})$. 
For instance, estimating the distances between states $\rho_i\in\mathcal{S}$ 
is used in quantum state certification protocols~\cite{Badescu2017a, yu2019quantum,fanizza2021testing}, which aim at giving statistical guarantees that a source of quantum states is truly iid by checking whether all $\rho_i$ are equal. Certifying that a source produces identical states in this way is much more resource efficient in comparison to a protocol based on quantum state tomography.

The simplest instance of relational property learning is that of estimating the overlap,  $|\braket{\psi}{\phi}|^2$, between two pure quantum states $\ket{\psi},\ket{\phi}$, given $n_1$ copies of $\ket{\psi}$ and $n_2$ copies of
$\ket{\phi}$. 
Overlap estimation plays an important role in, e.g., entanglement estimation \cite{Harrow2013}, quantum fingerprinting \cite{Buhrman2001}, or quantum machine learning through the HHL algorithm \cite{Harrow2009}.
In this case the operators $\xi_{y_1,y_2;y}(\ket{\psi},\ket{\phi})$ can be seen to be rank one, and the only admissible values for $y_1$ and $y_2$ are those associated to completely symmetric subspaces. All relevant information is thus contained in the probability distribution of $y$. Ref.~\cite{Badescu2017a} considered the optimal estimation of the Hilbert-Schmidt distance according to the worst-case figure of merit, whereas~\cite{Fanizza2020} considered optimal protocols for estimating the overlap, according to average figures of merit. For mixed states $\xi_{y_1,y_2;y}(\rho_1,\rho_2)$ are more complicated, and there are several inequivalent distance measures that one can estimate. In~\cite{Badescu2017a} an optimal algorithm for quantum state certification in trace distance was obtained, using the estimation of the squared Hilbert-Schmidt distance $D_{HS}^2(\rho_1-\rho_2)=\tr[(\rho_1-\rho_2)^2]$,  with sample complexity $\Theta(d/\epsilon^2)$.  The key to the construction is the observation that a good estimator for $D_{HS}^2$ can be obtained by noting that the latter is the trace of a polynomial function of the states.  For this type of functions one can obtain unbiased estimators from appropriate linear combinations of permutations. The corresponding measurement corresponds to a projective measurement on the 
various sectors $(y_1,y_2;y)$.  The ensuing joint probability distribution  concentrates on the spectra of $\rho_1,\rho_2,\,\bar \rho=\sum_{i=1,2} \frac {n_{i}}{n}\rho_i$, and $D_{HS}^2(\rho_1-\rho_2)$ depends only on the spectra of $\rho_1, \rho_2, \frac{\rho_1+\rho_2}{2}$. 

For collections of states, one can also ask about average distance measures of the form $\sum_{i=1}^lp_i h(\rho_i,\sum_{j=1}^n p_j\rho_j)$. Again, the joint distribution of $\vec y, y$ gives sufficient information for estimating classes of distance quantifiers that depend only on the spectra of convex combinations of the states, such as the Hilbert-Schmidt distance squared $D_{HS}^2$, or the Holevo quantity of $\mathcal{D}_Q$.
For other distance measures---such as the trace distance or the fidelity---an analysis of optimal learning protocols is lacking. Using the Hilbert-Schmidt distances one can also address the problem of identity certification of collections of quantum states, i.e., whether $\rho_i=\rho_j, \forall\, \rho_i,\rho_j\in\mathcal{S}$,  or the average trace distance  between the states is larger than $\epsilon>0$.  The sample complexity for this problem depends on the number of copies, $n_i$, of each state, $\rho_i$, we are 
given. If we are free to ask for copies of $\rho_i$ at will (the so-called query model~\cite{yu2019quantum}) then the sample complexity is still $\Theta(d/\epsilon^2)$.  If, on the other hand, $\rho_i$ are sampled from a distribution 
${p_i}$ (the so-called sampling model~\cite{fanizza2021testing}) then the sample complexity increases to $\Theta(\sqrt{l}d/\epsilon^2)$.

\section{Discussion}

Learning properties of quantum datasets will be an indispensable 
step of upcoming 
quantum-technological applications, be it as part of their normal operation or as a tool to certify that they work as intended. Indeed, the efficient certification of quantum states and processes has already been identified as a pressing need in near-term quantum applications~\cite{Eisert2020}, and protocols have been devised to test elementary properties assuming iid runs of quantum experiments. 
In this Perspective we have focused on quantum learning algorithms that go beyond the iid assumption and aim at more general types of properties of highly-structured quantum datasets, leveraging symmetries to handle the increased complexity of the problems. Conceivably, this kind of problems will become more and more 
relevant in the context of future 
quantum networks, where quantum data are the natural carriers of information~\cite{Wehner2018,Dunjko2020}.

While the works reviewed here are concerned with optimal algorithms, it is worth noting that these may require applying 
collective measurements over a (typically) large quantum dataset, which are generally hard to implement in practice. Still, analyzing optimal performance provides an ultimate benchmark of the learning task at hand against which we can test more feasible strategies, e.g., based on LOCC or sequential measurements \cite{MartinezVargas2021}.

Finally, let us mention that the approach to quantum learning covered here can also be formally extended to quantum objects beyond states, such as quantum channels or 
quantum processes. In these, the quantum ``dataset'' would consist in, e.g., a number of independent uses of an unknown device that we regard as a black box, and over which we may make certain structural assumptions. One example of such a problem would be detecting the presence and position of an anomaly in a sequence of allegedly identical unitary operations~\cite{Skotiniotis2018}, in a setting where the unitaries are not perfectly characterized.

\acknowledgments

We acknowledge financial support from the Spanish Agencia Estatal de Investigaci\'on, Grant No. PID2019-107609GB-I00 and Catalan Government for the project QuantumCAT  001-P-001644, co-financed by the European Regional Development Fund (FEDER). JC also acknowledges support from ICREA Academia award.

\bibliography{_EPL_perspective}


\end{document}